\documentclass[aps,twocolumn,prl,nofootinbib,letterpaper,nobibnotes,nobalancelastpage]{revtex4}

\usepackage{graphicx}
\usepackage{amssymb}

\newcommand{\mysum}[2]{\sum\limits_{#1}^{#2}}
\newcommand{\myint}[2]{\int\limits_{#1}^{#2}}

\newcommand{\mybpar}[1]{\left( #1 \right)}
\newcommand{\beq}{\begin{equation}}
\newcommand{\eeq}{\end{equation}}
\newcommand{\beqa}{\begin{eqnarray}}
\newcommand{\eeqa}{\end{eqnarray}}

\begin{document}
\title{Modeling inelastic phonon scattering in atomic- and molecular-wire junctions}
\author{Magnus \surname{Paulsson}}
\email{mpn@mic.dtu.dk}
\author{Thomas \surname{Frederiksen}}
\author{Mads \surname{Brandbyge}}

\affiliation{NanoDTU, MIC -- Department of Micro and Nanotechnology,
Technical University of Denmark, {\O}rsteds Plads, Bldg.~345E, DK-2800
    Lyngby, Denmark}
\date{\today}

\begin{abstract}
Computationally inexpensive approximations describing electron-phonon 
scattering in molecular-scale conductors are derived from the 
non-equilibrium Green's function method. The accuracy 
is demonstrated with a first principles calculation 
on an atomic gold wire. Quantitative agreement between the full 
non-equilibrium Green's function calculation and the newly derived 
expressions is obtained while simplifying the computational 
burden by several orders of magnitude. 
In addition, analytical models provide intuitive 
understanding of the conductance including non-equilibrium heating 
and provide a convenient way of parameterizing the physics. 
This is exemplified by fitting the expressions to the 
experimentally observed conductances through both an atomic gold 
wire and a hydrogen molecule. 
\end{abstract}

 \maketitle

%
%

The rapid evolution in electronics towards smaller
and faster devices will eventually reach the fundamental level set
by the atomistic structure of matter. Atomic-size conductors take
this development to the extreme of
miniaturization \cite{AgYeva.03.Quantumpropertiesof}, and understanding their
properties is an important problem in the emerging fields of
nanoelectronics and molecular electronics. One relevant aspect is
the study of the effects caused by atomic vibrations, since
inelastic scattering of traversing electrons and energy
dissipation play essential roles for device characteristics,
working conditions, and stability. Vibrational signals can also
be used to extract information about the detailed microscopic
configuration, which usually cannot be imaged simultaneously
with a transport measurement. Inelastic effects have in the recent
years been studied in a variety of nanoscale systems, e.g.,
single molecules on surfaces probed with the scanning tunneling
microscope (STM)  \cite{StReHo.98.Single-moleculevibrationalspectroscopy}, 
molecules in break junctions
\cite{SmNoUn.02.Measurementofconductance}, 
and metallic atomic wires \cite{AgUnRu.02.Onsetofenergy}. 

Theoretical descriptions of inelastic transport through 
small devices connected to metallic contacts include 
many-body theory in the Coulomb blockade regime 
\cite{BrFl.03.Vibrationalsidebandsand},
single-particle first-order perturbation 
approaches \cite{MoHoTo.03.Inelasticcurrent-voltagespectroscopy,ChZwDi.04.Inelasticcurrent-voltagecharacteristics},
i.e., ``Fermi's golden rule'' (FGR), as well as calculations to 
infinite order based on 
the self-consistent Born approximation (SCBA)
combined with non-equilibrium Green's functions
(NEGF) \cite{GaRaNi.04.Onlinewidths,FrBrLo.04.InelasticScatteringand,MiTiUe.03.Spectralfeaturesof}.
Our work is based on the SCBA, which in contrast to FGR
takes the many-particle nature of the problem into account.
However, the SCBA method is
computationally very demanding especially when used in combination
with first principles electronic structure methods. Moreover, the SCBA does not
yield simple formulas which can be used to extract information
from experimental data.

In this paper we develop methods which vastly simplify the  
SCBA approach. The main results are analytical formulas for the
current and power derived from a lowest order expansion (LOE) 
of the SCBA expressions. In particular, we show how 
first principles SCBA calculations 
on atomic gold wires can be accurately described by the LOE with 
minimal computational effort. Moreover, we derive 
compact analytical expressions using two simple models. 
These latter models are able to fit both the theoretical SCBA 
results as well as experiments using the electron-hole damping rate 
of the phonon as the central parameter \cite{PEPE.80}.

%
%

Phonon scattering is included in the SCBA method 
as self-energies to the electronic description. We use the undamped
phonon Green's functions to express these self-energies in the 
device subspace as \cite{Haug1996,PeCa.04.Atomistictheoryof}
\footnote{
  The polaron term \cite{HYHEDA.94} in the retarded self-energy in 
  Eq.~(\ref{eq.sigmaR}) has been neglected since it gives no ``signal''
  at the phonon energy. However, it gives rise to 
  two additional terms in the expression for the current
  (Eq.~(\ref{eq.current1})) proportional to $V$ and $V^2$ and does not 
  contribute to the power.
}:
\beqa
\mathbf{\Sigma}^{\lessgtr}_\mathrm{ph}(E) & = & 
  \mysum{\lambda}{} \mathbf{M}_\lambda 
     \left[(n_\lambda+1) \mathbf{G}^{\lessgtr}(E \pm \hbar \omega_\lambda)+ 
\nonumber \right. \\ & &
          + \left. 
             n_\lambda \mathbf{G}^{\lessgtr}(E \mp \hbar \omega_\lambda) 
          \right] \mathbf{M}_\lambda ,
\label{eq.sigma}
\\
\mathbf{\Sigma}^{r}_\mathrm{ph}(E) & = & \frac{1}{2} \mybpar{ 
   {\mathbf{\Sigma}^{>}_\mathrm{ph}(E)-\mathbf{\Sigma}^{<}_\mathrm{ph}(E)}}- 
\nonumber \\
   & &-  \frac{i}{2} {\cal H}\left\{{\mathbf{\Sigma}^{>}_\mathrm{ph}(E')-
     \mathbf{\Sigma}^{<}_\mathrm{ph}(E')} \right\}(E).
\label{eq.sigmaR}
\eeqa
Here, $\mathbf{M}_\lambda$ is the electron-phonon coupling matrix  
for phonon mode $\lambda$ occupied by $n_\lambda$ phonons with energy $\hbar \omega_\lambda$.
The lesser/\-greater self-energy matrices $\mathbf{\Sigma}^{\lessgtr}_\mathrm{ph}$ are 
given by two terms corresponding to absorption/emission of phonon quanta.
We furthermore assume that these self-energies can be used in non-equilibrium with a bias
dependent phonon occupation number $n_\lambda(V)$. 
The retarded self-energy can then be obtained from the greater/lesser parts 
using the Hilbert transform 
(${\cal{H}} \left\{ f(E') \right\}(E) = 1/\pi \, {\cal P} \int f(E')/(E-E') \, \mbox{d}E' $).

The computational difficulty of solving the SCBA equations 
stems from the coupling of Green's functions in energy. 
Calculations usually involve a numerical energy grid which 
has to be fine enough to resolve the 
low temperature structure of the Fermi function, while at the same time 
span a large energy range to cover phonon-energies, applied bias, 
and allow an accurate computation of the Hilbert transform 
which is nonlocal in energy. The current and power are then
computed as integrals over this energy grid 
\cite{Haug1996,PeCa.04.Atomistictheoryof,FrBrLo.04.InelasticScatteringand}.

These difficulties can be overcome if (i) the electron-phonon
coupling is weak, i.e., the probability for multi-phonon processes is low, 
and (ii) the density of states 
(DOS) of the contacts and the device is slowly varying over a 
few phonon-energies around the Fermi energy $E_F$, i.e., in the 
notation used below, 
$\mathbf{G}^r(E)\approx \mathbf{G}^r(E_F)$ and $\mathbf{\Gamma}_{1,2}(E) \approx \mathbf{\Gamma}_{1,2}(E_F)$. 
These approximations are valid for systems where (i) the 
electron spends a short time compared to the phonon scattering time in the 
device and (ii) the closest resonance energy ($E_\mathrm{res}$) is either 
far away from the Fermi energy  
($|E_\mathrm{res}-E_F|\gg \Gamma, eV$ and $\hbar \omega$) or the 
broadening by the contacts is large 
($\Gamma \gg eV, \hbar \omega$ and $|E_\mathrm{res}- E_F|$). 
The expressions for the current and power 
\cite{Haug1996,PeCa.04.Atomistictheoryof,FrBrLo.04.InelasticScatteringand} 
can then be 
expanded to lowest order (second) in the electron-phonon coupling 
and the integration over energy performed analytically. 
The power dissipated into the phonon system $P^\mathrm{LOE}$ can, 
after lengthy derivations, be written:
\begin{widetext}
\beqa
P^\mathrm{LOE}&=& \mysum{\lambda}{}
  \frac{\mybpar{\hbar \omega_\lambda}^2}{\pi \hbar} 
    \left( n_{B}(\hbar \omega_\lambda)  -n_\lambda \right)
    \, \mathrm{Tr}\left[ \mathbf{M}_\lambda \mathbf{A} \mathbf{M}_\lambda \mathbf{A} \right] 
   +  
    {\cal P}(V,\hbar \omega_\lambda,T)
    \, 
    \mathrm{Tr}\left[\mathbf{M}_\lambda \mathbf{G} \mathbf{\Gamma}_1 \mathbf{G}^\dag \mathbf{M}_\lambda 
        \mathbf{G} \mathbf{\Gamma}_2 \mathbf{G}^\dag \right] , 
  \label{eq.power} \\ 
{\cal P}&=&  
   \frac{\hbar \omega}{\pi \hbar}  \frac{
    \left( \cosh \left( \frac{e V}{kT} \right)-1\right)   \coth \left( \frac{\hbar \omega}{2 kT} \right) \hbar \omega
      -{e V} \sinh \left( \frac{e V}{kT} \right)
   }{
     \cosh \left( \frac{\hbar \omega}{kT} \right)-\cosh \left( \frac{e V}{kT} \right)} , \label{eq.powerfunc}
\eeqa
\end{widetext}
where $n_{B}$ is the Bose-Einstein distribution,
which appears naturally from the integration of the Fermi
functions of the electrons in the contacts. Here, 
$\mathbf{G}=\mathbf{G}^r(E_F)$, $\mathbf{\Gamma}_{1,2}=\mathbf{\Gamma}_{1,2}(E_F)$, and 
$\mathbf{A}=i(\mathbf{G}-\mathbf{G}^\dagger)$ are the non-interacting, i.e.,  without phonon interactions, 
retarded Green's function, 
the broadening by the contacts, and spectral function at $E_F$, respectively. 

From Eq.~(\ref{eq.power}) we see that the power can be decomposed into 
terms corresponding to the individual phonon modes. 
We also note that the first term describes the power
balance between the electron and phonon systems (at zero bias) with 
an electron-hole damping rate 
$\gamma^\lambda_\mathrm{eh}=\omega_\lambda / \pi  \, 
\mathrm{Tr}\left[ \mathbf{M}_\lambda \mathbf{A} \mathbf{M}_\lambda \mathbf{A} \right]$ and is 
in fact equivalent to the FGR expression \cite{PEPE.80,HETU.92}. The second term is 
even in bias and 
gives the phonon absorption/emission at non-equilibrium;
it is negligible at low bias ($e V\ll\hbar \omega$), turns on at the 
phonon energy and becomes linear in voltage at high bias 
($e V\gg\hbar \omega$).

Using the same approximations, the current through the 
device $I^\mathrm{LOE}$ is given by \cite{Viljas}:
\begin{widetext}
\beqa
 I^\mathrm{LOE} &=&   \frac{e^2 V}{\pi \hbar} \mathrm{Tr}
\left[\mathbf{G} \mathbf{\Gamma}_2 \mathbf{G}^\dag \mathbf{\Gamma}_1\right]  \nonumber \\ & & + 
\mysum{\lambda}{} {\cal I}^\mathrm{Sym}(V,\hbar \omega_\lambda, T, n_\lambda)  \,
   \mathrm{Tr}\left[
     \mathbf{G}^\dag \mathbf{\Gamma}_1 \mathbf{G} \left\{ 
                        \mathbf{M}_\lambda \mathbf{G} \mathbf{\Gamma}_2 \mathbf{G}^\dag \mathbf{M}_\lambda +
       \frac{i}{2}\left( \mathbf{\Gamma}_2 \mathbf{G}^\dag \mathbf{M}_\lambda \mathbf{A} \mathbf{M}_\lambda - \mathrm{h.c.}\right)
       \right\}
   \right]  \nonumber \\ 
 & & + 
\mysum{\lambda}{} {\cal I}^\mathrm{Asym}(V,\hbar \omega_\lambda, T)  \,
  \mathrm{Tr}\left[
         \mathbf{G}^\dag \mathbf{\Gamma}_1 \mathbf{G} \left\{
           \mathbf{\Gamma}_2 \mathbf{G}^\dag \mathbf{M}_\lambda \mathbf{G} \left( \mathbf{\Gamma}_2-\mathbf{\Gamma}_1 \right) \mathbf{G}^\dag \mathbf{M}_\lambda + 
           \mathrm{h.c.}
           \right\}
    \right] ,
 \label{eq.current1} \\
{\cal I }^\mathrm{Sym} & = & 
      \frac{e }{\pi \hbar} \mybpar{{2 e V} n_\lambda+
      \frac{\hbar \omega_\lambda-{e V}}{e^{\frac{\hbar \omega_\lambda-e V}{kT}}-1}-
      \frac{{\hbar \omega_\lambda}+{e V}}{e^{\frac{\hbar \omega_\lambda+e V}{kT}}-1}} ,
\label{eq.currentNormal} \\
 {\cal I}^{\mathrm{Asym}}&=&
    \frac{e}{2 \pi \hbar}\myint{-\infty}{\infty}  \left[n_{F}(E) -n_{F}(E-e V)\right] \,
    {\cal H} \left\{{n_{F}(E'+\hbar \omega_\lambda)-n_{F}(E'-\hbar \omega_\lambda)}\right\}(E)\, \, dE ,
   \label{eq.currentHilbert}
\eeqa
\end{widetext}
where $n_{F}$ is the Fermi function, the bias is defined via, 
$e V=\mu_2-\mu_1$, 
and the conductance quantum $G_0=e^2/\pi \hbar$ appears naturally. 
In contrast to the 1st order Born approximation, these expressions are 
current conserving like SCBA.

\begin{figure}[tb!]
    \begin{center}
 \includegraphics[width=0.9 \columnwidth,angle=0]{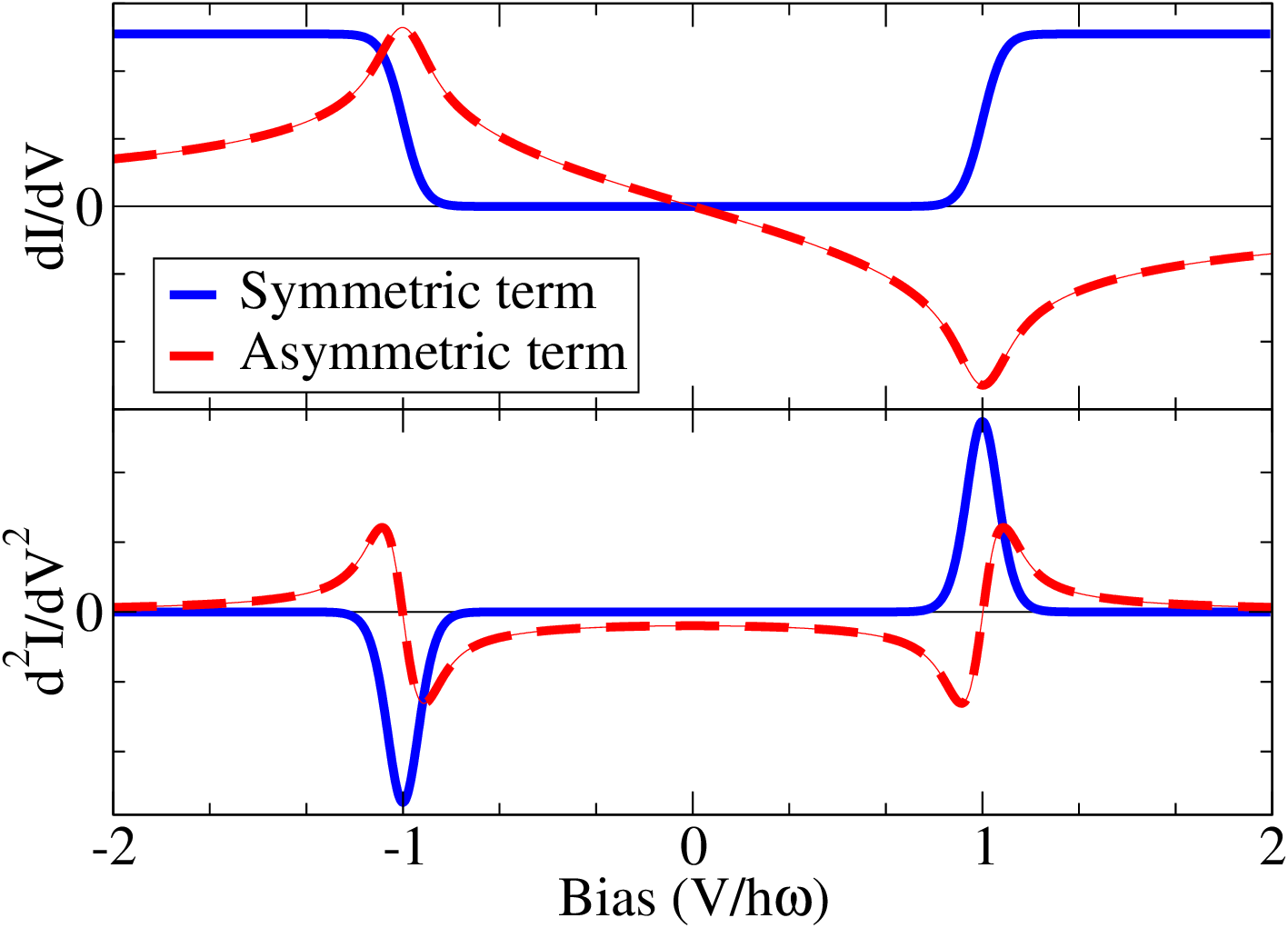}
  \caption{Universal functions (Eqs.~(\ref{eq.currentNormal}) and (\ref{eq.currentHilbert}))
giving the phonon contribution to the current. 
The differential conductance $dI/dV$ and the second derivative signals is shown for one 
phonon mode with the bias in units of the phonon energy at a temperature $kT=0.025 \, \hbar \omega$.
For the symmetric term, the FWHM of the second derivative peak 
is approximately $5.4 \, kT$ \cite{HA.77.INELASTICELECTRON-TUNNELING}.  
}
  \label{fig.functionplot}
    \end{center}
\end{figure}


The current expression retains the structure of the Landauer expression (first term of 
Eq.~(\ref{eq.current1})) and gives correction terms for each phonon mode. 
The phonon terms can in turn be divided into a ``symmetric'' term 
${\cal I}^{\mathrm{Sym}}$ where the differential conductance
$dI/dV$ is even in bias,
and an ``asymmetric'' term containing the Hilbert transform 
${\cal I}^\mathrm{Asym}$ 
yielding an odd contribution.
We note the simple factorization into terms depending on the
electronic structure at $E_F$ and \emph{
universal} functions ${\cal I}^{\mathrm{Sym}}$ and ${\cal I}^\mathrm{Asym}$ 
which yield the line-shape of the inelastic
signals in the $I-V$, see Fig.~\ref{fig.functionplot}.
Whether the conductance increases or decreases due to phonon scattering depends
on the sign of the traces in Eq.~(\ref{eq.current1}) and will be discussed further 
below. Examination of the ``asymmetric'' term in Eq.~(\ref{eq.current1}) shows 
that it is zero for symmetric systems.
Although experimentally measured conductances contain asymmetric signals, 
the size of these signal is usually small in the published curves. At
present it is unclear if they are caused by phonons or other effects.

As we have shown previously heating of the phonon system 
should be considered \cite{FrBrLo.04.InelasticScatteringand} 
which makes the number of phonons $n_\lambda$ bias dependent.
The simplest way to include non-equilibrium heating is to write down 
a rate equation, including an external damping rate $\gamma^\lambda_{\mathrm{d} }$ of the phonons:
\beq
\dot n_\lambda = \frac{P^\mathrm{LOE}_\lambda}{\hbar \omega} + 
   \gamma^\lambda_{\mathrm{d}} \mybpar{n_B(\hbar \omega_\lambda)-n_\lambda} ,
\label{eq.heating}
\eeq
where $P^\mathrm{LOE}_\lambda$ is the power dissipated into the 
individual phonon modes 
\footnote{
For weak electron-phonon interaction, the division of power into the individual phonon modes 
is straightforward from Eq.~(\ref{eq.power}).
}. 
The steady state occupation $n_\lambda$ is easily found. Substituting the result into Eqs.~(\ref{eq.current1})-(\ref{eq.currentHilbert}) 
gives a computationally simple but 
powerful formula for the current through the device including heating of 
the phonon system.

\begin{figure}[tb!]
\begin{center}
 \includegraphics[width=0.9 \columnwidth,angle=0]{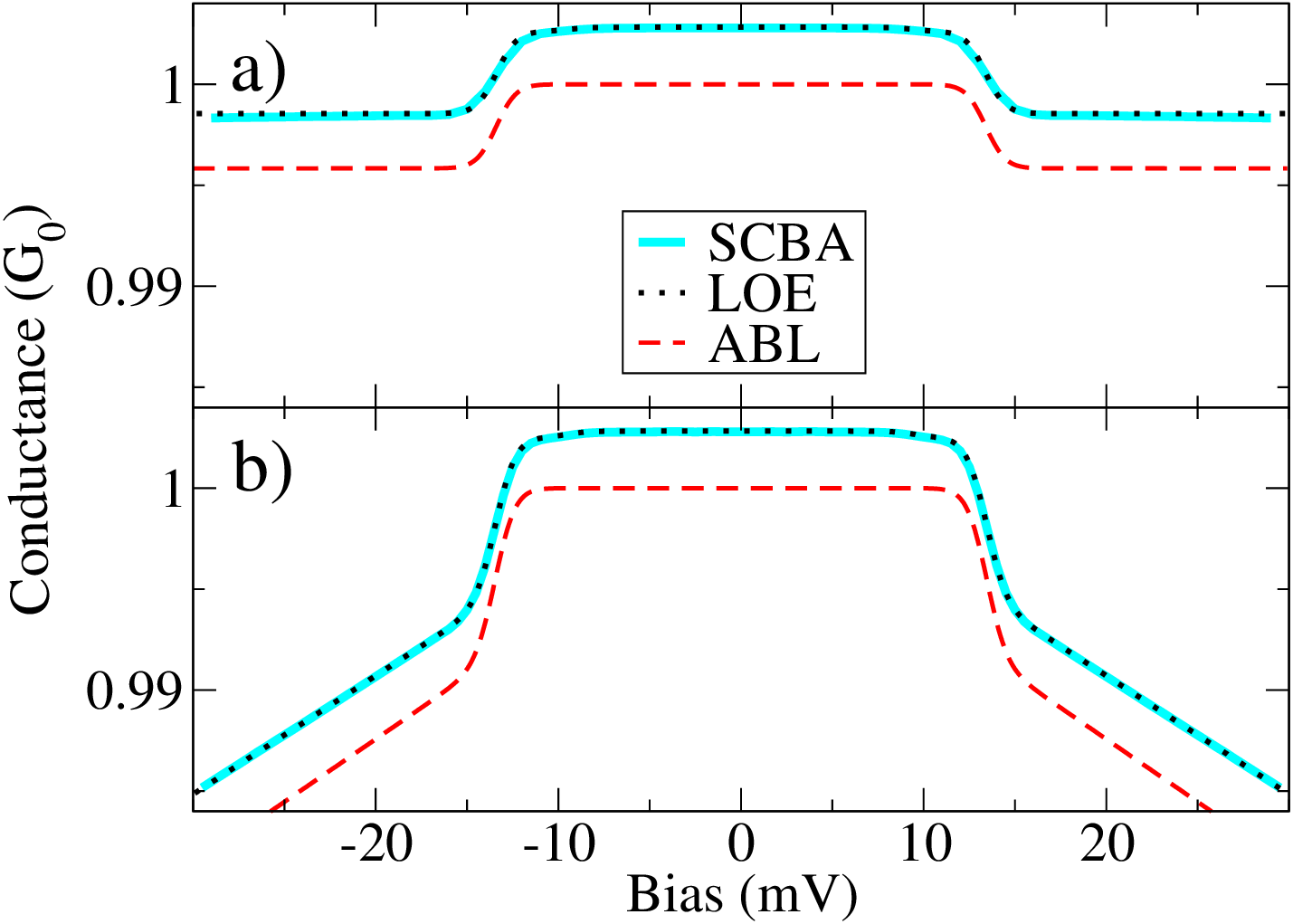}
  \caption{
Comparison between the SCBA results and the LOE 
expressions (Eq.~(\ref{eq.current1})) without heating a) and with heating b) 
($\gamma_\mathrm{d}=0$) at $T=4.2 \, \mathrm{K}$. 
The parameters for the ABL model 
(Eq.~(\ref{eq.currentau})) were extracted directly from the DFT calculations, 
$\gamma_\mathrm{eh}=5.4 \times 10^{10} \, \mathrm{s}^{-1}$ and 
$\hbar \omega=13.4 \, \mathrm{meV}$.
}
  \label{fig.auchain}
\end{center}
\end{figure}


To judge the accuracy of the LOE approach, we compare the LOE results 
to the full SCBA solution for a four atom gold wire, see 
Fig.~\ref{fig.auchain}. The SCBA calculation was performed as described 
previously \cite{FrBrLo.04.InelasticScatteringand}, where the Hamiltonian, phonon 
modes, and electron-phonon interaction were obtained from density 
functional calculations (DFT). 
The excellent agreement between the full SCBA and the LOE expression can 
be understood by noting that the DOS of a gold wire is slowly changing 
over an energy range much greater than the phonon energies. In addition,
the electrons only spend a small time in the wire
\cite{MoHoTo.03.Inelasticcurrent-voltagespectroscopy} 
compared to the electron-hole damping rate.
Importantly, the LOE conductance calculations were performed in less 
than a minute on a regular PC, compared to several hours for the 
SCBA calculations. The LOE approach thus opens up the possibility to 
study inelastic scattering with first principles methods for large 
systems, e.g., organic molecules.

To gain further insight into the expressions presented above, we consider
a single electronic site with symmetric 
contacts $\Gamma=\Gamma_{1}=\Gamma_{2}$ coupled to one phonon mode. 
Introducing the transmission probability 
$\tau = \vert G\vert^2 \Gamma^2$ and the electron-hole damping rate 
$\gamma_\mathrm{eh}=4 (\omega/ \pi) \, M^2 \tau^2 / \Gamma^2$ 
, we obtain:
\beqa
P^\mathrm{LOE}_\mathrm{one} & =&    \gamma_\mathrm{eh} \, \hbar \omega 
\left(n_{B}(\hbar \omega)  -n \right)  
  \label{eq.powerone} 
+\frac{\gamma_\mathrm{eh}}{4} \, \frac{\pi \hbar}{\hbar \omega}{\cal P} ,
 \\
I^\mathrm{LOE}_\mathrm{one} &=& \frac{e^2}{\pi\hbar} \tau V + 
e \gamma_\mathrm{eh} \frac{1-2 \tau}{4 }  
\frac{\pi \hbar}{e \, \hbar \omega}{\cal I}^\mathrm{Sym} .
\label{eq.currentone} 
\eeqa
We note that, from the term $1-2 \tau$ in Eq.~(\ref{eq.currentone}),
the conductance will increase due to phonon scattering for low 
conductance systems ($\tau < 1/2$) and decrease for highly conducting 
systems ($\tau > 1/2$).  The LOE approach directly provides the sign 
of the conductance change in contrast to FGR approaches 
where this requires careful considerations 
\cite{MoHoTo.03.Inelasticcurrent-voltagespectroscopy,ChZwDi.04.Inelasticcurrent-voltagecharacteristics}. 

\begin{figure}[tb!]
    \begin{center}
 \includegraphics[width=0.9 \columnwidth,angle=0]{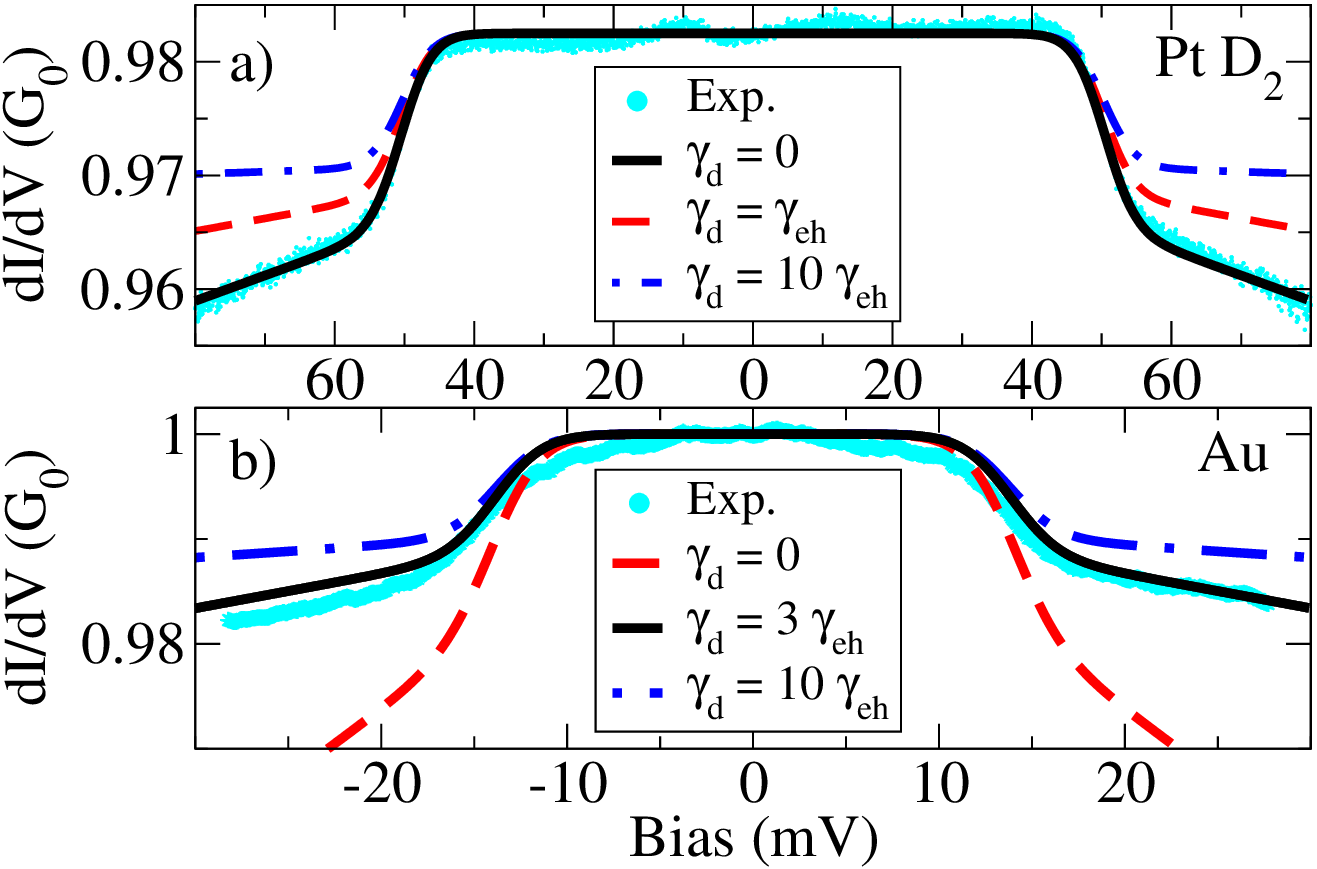}
  \caption{a) Single level model 
(Eqs.~(\ref{eq.powerone}-\ref{eq.currentone})) fitted 
to the experimentally measured conductance through a Deuterium molecule 
\cite{DjThUn.05.Stretchingdependenceof}. The parameters used for the fit are 
$\hbar \omega = 50 \, \mathrm{meV}$, 
$\tau=0.9825$, $\gamma_\mathrm{eh}=1.1 \times 10^{12} \, \mathrm{s}^{-1}$, and
$T=17 \, \mathrm{K}$.
b) The ABL model (Eqs.~(\ref{eq.powerau}-\ref{eq.currentau})) fitted to the measured 
conductance through 
an atomic gold wire (experimental data from Ref.~\cite{AgUnRu.02.Onsetofenergy}). 
The fit reveals the following parameters, 
$\hbar \omega =13.8 \, \mathrm{meV}$, $T=10 \, \mathrm{K}$, 
$\gamma_\mathrm{eh}=12 \times 10^{10} \, \mathrm{s}^{-1}$, 
and $\gamma_\mathrm{d}=3 \gamma_\mathrm{eh}$.
}
  \label{fig.hydrogen}
    \end{center}
\end{figure}


The conductance through a single hydrogen molecule has been measured using a 
platinum break junction setup \cite{SmNoUn.02.Measurementofconductance,
DjThUn.05.Stretchingdependenceof}. Because the
elastic current is carried through a single molecular orbital 
\cite{DjThUn.05.Stretchingdependenceof},
the single level model fits the experiment very well, see Fig.~\ref{fig.hydrogen}a. 
The best fit is obtained using a negligible external damping of the 
phonon mode ($\gamma_\mathrm{d} \ll \gamma_\mathrm{eh}$) which can be understood 
physically from the mass difference between the hydrogen molecule and the platinum atoms 
of the  break junction. We also note that  both the size of the 
conductance step and the conductance slope (caused by heating) is fitted with 
only one parameter, the electron-hole damping rate $\gamma_\mathrm{eh}$.

The electronic structure of atomic gold chains are qualitatively different 
from the one level model.
However, it is relatively straightforward to 
derive an alternating bond length (ABL) model. 
Inserting the electron-phonon matrix for an ABL 
phonon mode \cite{FrBrLo.04.InelasticScatteringand} and 
using the Green's function for a  
half filled perfectly transmitting 1-D chain we obtain:
\beqa
P^\mathrm{LOE}_\mathrm{ABL} & =&   \gamma_\mathrm{eh} \, \hbar \omega 
\left[n_{B}(\hbar \omega)  -n \right]  
  \label{eq.powerau} 
+\frac{\gamma_\mathrm{eh}}{2} \, \frac{\pi \hbar}{\hbar \omega}{\cal P} ,
 \\
I^\mathrm{LOE}_\mathrm{ABL} &=& \frac{e^2}{\pi\hbar}  V - 
\frac{ e \gamma_\mathrm{eh}}{2}   
\frac{\pi \hbar}{e \, \hbar \omega }{\cal I}^\mathrm{Sym} ,
\label{eq.currentau} 
\eeqa
where the only difference to the one-level model is that $\tau=1$
(perfect transmission) and a factor of two caused by the absence of forward 
scattering from an ABL mode 
(the one-level model has an equal amount of forward and back scattering). 
The ABL model is shown in Fig.~\ref{fig.auchain}, with the $\gamma_\mathrm{eh}$ 
damping rate calculated directly from the DFT model. 
The main difference compared to the SCBA/LOE results is the assumption of 
perfect transmission through the chain. 
Fitting the ABL model to experimental data \cite{AgUnRu.02.Onsetofenergy} gives
the very satisfactory fit shown in Fig.~\ref{fig.hydrogen}b. We briefly
note that the external damping $\gamma_\mathrm{d}=3 \, \gamma_\mathrm{eh}$ 
is not negligible in contrast to the hydrogen case.
In this paper we have used sharp phonon energies, c.f., Eq.~(\ref{eq.sigmaR}). 
However, if the phonon spectral function is known, it is possible to introduce 
broadening directly into Eqs.~(\ref{eq.power})-(\ref{eq.currentHilbert}) 
from a finite phonon lifetime.

We have derived simple and accurate approximations to describe the 
effect of phonon scattering on the conductance through nanoscale 
conductors. The approximate expressions greatly reduce the computational 
effort, compared to solving the SCBA equations. 
In addition, simple models were derived which provide insight and are 
suitable to fit experimental data.

\begin{acknowledgments}
The authors are grateful to D. Djukic, J. M. van Ruitenbeek, and 
N. Agra\"it for helpful discussions regarding their experimental 
work. This work, as part of the European Science Foundation EUROCORES 
Programme SASMEC, was supported by funds from the SNF and the EC 6th 
Framework Programme. Computational resources were provided by DCSC.
\end{acknowledgments}

\bibliography{Manus}
\bibliographystyle{apsrev}

\end{document}